\documentclass[journal]{IEEEtran}

\usepackage{xcolor,soul,framed} 
\usepackage{multirow}
\colorlet{shadecolor}{yellow}
\usepackage[pdftex]{graphicx}
\graphicspath{{../pdf/}{../jpeg/}}
\DeclareGraphicsExtensions{.pdf,.jpeg,.png}

\usepackage[cmex10]{amsmath}
\usepackage{array}
\usepackage{mdwmath}
\usepackage{mdwtab}
\usepackage{eqparbox}
\usepackage{url}
\usepackage{subfigure}
\usepackage[justification=centering]{caption}

\begin{document}
\bstctlcite{IEEEexample:BSTcontrol}
    \title{Tire Force Estimation in Intelligent Tires Using Machine Learning}

\author{
		Nan Xu,~Hassan Askari,~Yanjun Huang,~Jianfeng Zhou and Amir Khajepour
	
\thanks{The Copyright of this work has been transferred to IEEE. The full citation is: N. Xu, H. Askari, Y. Huang, J. Zhou and A. Khajepour, "Tire Force Estimation in Intelligent Tires Using Machine Learning," in IEEE Transactions on Intelligent Transportation Systems, doi: 10.1109/TITS.2020.3038155. And the link of this work is https://ieeexplore.ieee.org/document/9284471. This research is supported by National Natural Science Foundation of China (Grant Nos.51875236 and 61790561), China Automobile Industry Innovation and Development Joint Fund (Grant Nos. U1664257 and U1864206)}
\thanks{N. Xu and J. Zhou are with the State Key Laboratory of Automotive Simulation and Control, Jilin University, Changchun, Jilin, 130025, China and N. Xu is also with the Department of Mechanical and Mechatronics Engineering, University of Waterloo, ON. N2L3G1, Canada, e-mail: (xunan@jlu.edu.cn).}
\thanks{Yanjun Huang is  at the School of Automotive Studies, Tongji University, Shanghai, 201804, China, e-mail: (yanjun\_huang@tongji.edu.cn).}
\thanks{H. Askari and A.Khajepour are  at the Department of Mechanical and Mechatronics Engineering, University of Waterloo, ON. N2L3G1, Canada e-mail: ( h2askari@uwaterloo.ca, and a.khajepour@uwaterloo.ca) }}

\maketitle

\begin{abstract}
The concept of intelligent tires has drawn attention of researchers in the areas of autonomous driving,~advanced vehicle control, and artificial intelligence.~The focus of this paper is on intelligent tires and the application of  machine learning techniques to tire force estimation.~We present an intelligent tire system with a tri-axial acceleration sensor, which is installed onto the inner liner of the tire, and Neural Network techniques  for real-time processing of the sensor data.~The accelerometer is capable of measuring the acceleration in x,y, and z directions.~When the accelerometer enters the tire contact patch,~it starts generating signals until it fully leaves it.~Simultaneously,~by using MTS Flat-Trac test platform, tire actual forces are measured.~Signals generated by the accelerometer and MTS Flat-Trac testing system are used for training three different machine learning techniques with the purpose of online prediction of tire forces.~It is shown that the developed intelligent tire  in conjunction with machine learning is effective in accurate prediction of tire forces under different driving conditions.~The results presented in this work will open a new avenue of research in the area of intelligent tires, vehicle systems,~and tire force estimation.

\end{abstract}


%
\IEEEpeerreviewmaketitle


\section{Introduction}
\IEEEPARstart{T}{ire} and road interaction is the main source of force generation in terms of vehicle dynamics, and online estimating of tire forces are important to vehicle safety.~Thus, it is highly desirable to have an online measurement or estimation of tire forces \cite{cheli2011cyber,askari2019tire,rezaeian2014novel}.~The need of online tire measurement system is more pronounced for autonomous vehicles and all vehicles when they experience harsh maneuver in which estimation techniques fail to accurately report tire  forces, moments, and slip angles.~An intelligent tire system is defined as a system that can intelligently extract and send valuable information about tire and road conditions to vehicles’ Electronic Control Unit (ECU).~By using different communication protocols including 5G network \cite{kousaridas20205g}, intelligent tires can be actualized in the near future to make vehicles safer and fully connected.~Thus,~it is paramount to develop novel sensing systems along with novel machine learning techniques with the potential of real-world applications and implementation in tires \cite{askari2019embedded}.

One crucial step for the actualization of intelligent tires is to accurately predict tire forces.~Research in the area of tire forces spans from a simple formula to complicated finite element model and direct measurement techniques with extra sensory devices added to tires.~Considering theoretical aspects, tire models have been developed based on their frequency range including steady state, transient and high frequency tire models \cite{pacejka2005tire,chen2015research}.~Depending on the working conditions, tire models are also classified as pure and combined slip models.~Tire models include but not limited to  Magic Formula \cite{pacejka2005tire}, average lumped  LuGre \cite{canudas2003dynamic, velenis2002extension},~UniTire \cite{guo2005unitire},~Kamm circle \cite{kiencke2000automotive},~Nicolas-Comstock \cite{brach2000modeling} and Dugoff \cite{ghandour2010tire}.~These models can somehow predict tire forces; however, their parameters should be tuned experimentally, which makes their practical applications harder as the aging and wear of tires plus complicated driving conditions affect those parameters.

Different estimation techniques have been also developed and utilized by researchers to estimate tire forces.~These techniques are used to find vertical,~longitudinal and lateral tire forces \cite{cho2009estimation,doumiati2009lateral,wilkin2006use,viehweger2020vehicle}.~Viehweger et al. \cite{viehweger2020vehicle} summarized four model-based methods to estimate vehicle states and tire forces in all three directions in their recently published article.~The most well-known techniques used for tire forces estimation are linear \cite{farhat2017tire}, RLS-based \cite{nam2012estimation},~sliding mode \cite{baffet2009estimation}, nonlinear \cite{hashemi2017corner},~unknown input observers \cite{wang2012longitudinal} and Kalman-based techniques \cite{rezaeian2014novel,cheng2019simultaneous}.~For the estimation of vertical forces, generally, the longitudinal and lateral load transfers plus the static load of each tire are used. For example, Doumiati et al. \cite{doumiati2009lateral,doumiati2010observers} modeled tire vertical forces with the coupling of the longitudinal and lateral loads transfers, and proposed an algorithm to estimate lateral load transfer and vertical forces based on the Kalman filter. In another work, Cho et al. \cite{cho2009estimation} studied tire vertical forces by adding the longitudinal and lateral load transfers to static loads.~Very recently, Cordeiro et al. \cite{cordeiro2019estimation} used a delayed interconnected cascade-observer structure along with Extended Kalman Filter~(EKF) and Unscented Kalman Filter (UKF) to find tire forces including the vertical one.~The implementation of delayed interconnections addresses the challenges related to the mutual dependence in cascade estimator.

Estimation of tire longitudinal force is very important to determine the performance behavior of vehicles.~Different techniques have been used to find tire longitudinal forces.~For example, EKF method was adopted to calculate the tire longitudinal force as state estimates, which can avoid complex tire model parameters \cite{wilkin2006use}.~Random-walking model was used by Cho et al. \cite{cho2009estimation} and Rajamani et al. \cite{rajamani2011algorithms} to estimate tire longitudinal forces.~One of the issues of using  random walk models in tire longitudinal forces estimation is related to  observers as the model dynamic information is not fully explored by them in the state correction process \cite{cordeiro2019estimation}.~A few other researchers adopted sliding mode observers to identify tire longitudinal force.

Estimation of tire lateral forces is the most challenging one in comparison with the other two forces because of the  observability conditions \cite{hashemi2017corner}.~Several researchers have focused on the estimation of tire lateral forces.~For instance, by utilizing a random-walk Kalman filter, tire lateral forces were estimated in \cite{doumiati2010observers}.~Doumiati et al. \cite{doumiati2010onboard} exploited EKF and UKF observers to identify tire lateral force and side-slip angles.~In another study, according to a nonlinear vehicle dynamics model, the interacting multiple model-unscented Kalman filter (IMM-UKF) and the interacting multiple model-extended Kalman filter (IMM-EKF) were used to estimate tire lateral forces \cite{jin2015estimation}.~Assuming the time derivative of the lateral force is proportional to the roll rate, K. Huh \cite{huh2001active} estimated the lateral tire force of each wheel using a four degrees of freedom vehicle model.~However, several estimation works have been performed to find tire forces,~they rely heavily on tire and vehicle models with different complexity levels and are still insufficient and unreliable when vehicles experience harsh maneuver.~In addition, such indirect methods introduce additional uncertainties to the control procedure and may bring inaccuracies such as integration errors or time lags, which might be a disaster for the driver and driver assist systems when facing an emergency.~Accurate estimation of tire forces is very important because the vehicle performance and safety highly depends on the performance of control algorithms.~The control components of vehicles determine their control strategy and parameters according to the estimated tire forces and moments. Accordingly, several researchers have focused on the application of intelligent tires from the perspective of vehicle control, because intelligent tire technique is proved to provide more accurate tire variables of interest by processing measurements of tire deformations or strains using sensors installed inside the tire.~In addition, it is expected that control algorithms associated with intelligent tire technology can be more effective because uncertainties in the observers will be disappeared \cite{Lee2017review}.~Accordingly,~many  sensory devices have been developed for locating them inside the tire/hub to provide useful and online information from tire and road interaction \cite{askari2019tire,askari2017triboelectric}. A Load-Sensing Hub Bearings (LSB) unit was developed with using strain gauges to relate the tire forces and moments to the bearing deflections \cite{Den2013LSB}.~Several kinds of sensing systems were designed for the development of intelligent tire to directly measure tire dynamics parameters \cite{hu2011nanogenerator, wang2015nonlinear,li2015modeling}.~The data extracted using the direct measurement of tires are usually fused with estimation techniques or tire models to give a better prediction from tire forces.

One of the main challenges of developing intelligent tires is related to accurate interpretation of the generated data from the sensors during different driving scenarios.~Traditional analytical methods try to establish the relationship by physical model, but it is difficult to understand and model the tire behavior under complex driving conditions \cite{Lee2017review}.~This shortcoming in the areas of tire forces estimation and intelligent tires have prompted us to develop a novel intelligent tire using accelerometer and machine learning techniques.~Different machine learning methods have been used in the automated driving, tire modeling and parameter estimations for vehicle dynamics application.~The feasibility of the neural network used for tire modeling is investigated by Kim with feedforward back propagation algorithm \cite{Kim1995NN} and Boada with the recursive lazy learning method \cite{Boada2011NN}, and the advantages under a complex and wide range of tire environments were shown. Cramer trained a decision tree-based model to build the relationship of vehicle dynamics objective measurements and subjective assessment scores from professional drivers \cite{Cramer2017DT}. Roychowdhury applied a convolutional neural network (CNN) to learn region-specific features to classify road surface condition using front-camera images \cite{Roychowdhury2018CNN}.~To develop an affordable real-time snow detection system, Khan used SHRP2 naturalistic driving study video data to train the developed system. Two texture-based image features and three classification algorithms, namely support vector machine (SVM), k-nearest neighbor (K-NN), and random forest (RF), were used to classify the image groups \cite{Khan2019RF}.~Authors from Porsche AG and the University of Duisburg-Essen \cite{Graber2019RNN} investigated the use of a recurrent neural network (RNN) with gated recurrent units for side-slip angle estimation.~To further increase the performance and robustness of the estimation, a simplified vehicle model is incorporated.~Nearly 6 million data points on dry, wet, and snowy road surface conditions were used.~Their results show a great estimation quality in all covered situations.

Generally, the main machine learning methods used in automotive industry, especially for tire/vehicle parameters estimation, contain the family of decision tree algorithms (Decision Tree, Random Forest, Gradient boosting machines, etc.) and the family of neural networks (Artificial Neural Networks, Convolutional Neural Networks, Recurrent Neural Networks, etc.). An accepted conclusion in machine learning field is that no single algorithm performs the best across all possible scenarios. Therefore, in this study, three commonly used algorithms, Neural Network, Random Forest and Recurrent Neural Network, are implemented for the sake of comparison. Neural Networks usually consist of an input, an output layer, and one or more hidden layers that converts the input into something that the output layer can use \cite{Hecht1992ANN}.~A Neural Network can process all types of data coded in the numeric form.~Neural Networks can be formulated deeper via increasing the number of hidden layers.~The more hidden layers the more complex the representation of an application can be.~Random Forest algorithm uses and merges the decisions of several decision trees to reach an answer, which basically describes the average of all existing decisions trees \cite{ho1998random}. In this way, it compensates for possible errors of single trees in the forest such that the model is less prone to produce results further away from the real values.~In many practical applications, the Random Forest have been proven to be a very potent learning approach.~However, when used in regression problems, Random Forest  is subjected to the limitation that they cannot go beyond the range of values of the target variable utilized in training \cite{Ellis2016RF}.~RNN is considered as a class of neural networks, which can capture temporal dependencies and thus are well appropriate in processing sequence data to make predictions \cite{dupond2019thorough}.~Generally, this technique fits well to natural language processing and speech recognition applications.~A comparative results are provided, and finally the neural network with Rprop algorithm is adopted as the most effective approach for tire forces estimation because of its effective performance in terms of accuracy and easy implementation.

As illustrated, in this study, the intelligent tire is developed by attaching a MEMS-based accelerometer to its inner liner. The accelerometer is capable of measuring the acceleration in x, y, and z directions.~The accelerations measured by the accelerometer are used for training three different  machine learning techniques described above with the aim of finding the interaction forces between the tire and road.~This is the first work in which tire forces are predicted in all directions using machine learning in an intelligent tire system.~In the next sections, we first present the experimental setup used for online measurement of tire forces and the intelligent tire system. Then, the machine learning algorithms used in this work, are briefly presented. In addition, it is shown how the collected data is used for the training purposes. The very last section of the paper shows the potency of trained neural network for the prediction of tire forces.~We first start with the experimental setup as it will be presented in the next section.
\section{Experimental Setup}\label{expriment}
\begin{figure}[ht]
  \begin{center}
  \includegraphics[width=3.5in]{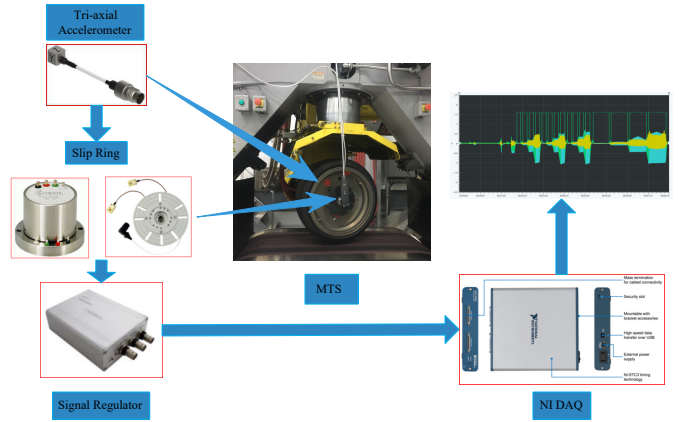}
  \caption{The intelligent tire testing system}\label{Figure_1}
  \end{center}
\end{figure}

\begin{figure}
    {
        \includegraphics[width=1in]{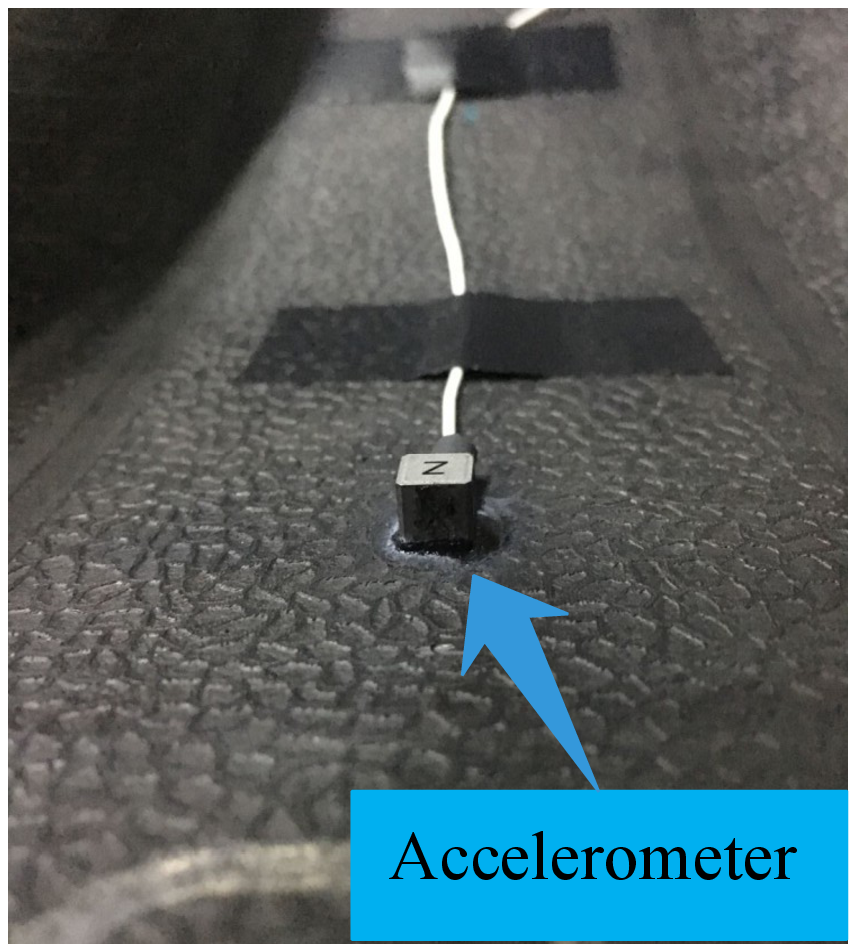}{a}
        \label{fig:first_sub}
    }
    \centering
    {
        \includegraphics[width=1in]{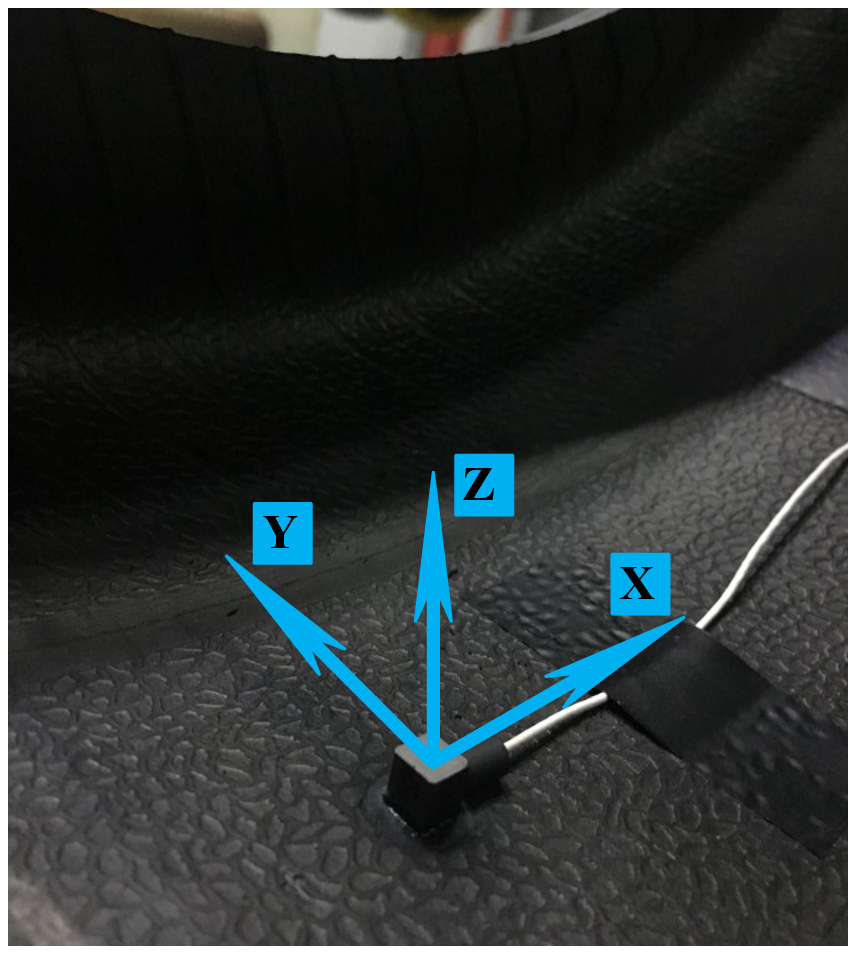}{b}
        \label{fig:second_sub}
    }
    \caption{(a) Accelerometer attached to the inner liner of the tire, (b)  its coordinate system}
    \label{Figure_2}
\end{figure}

\begin{figure}
    {
        \includegraphics[width=1in]{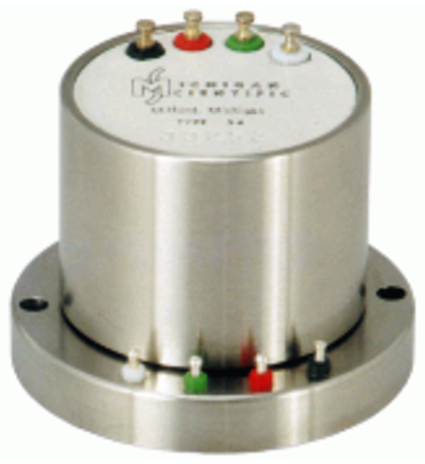}{a}
        \label{fig:first_sub}
    }
    \centering
    {
        \includegraphics[width=2.0in]{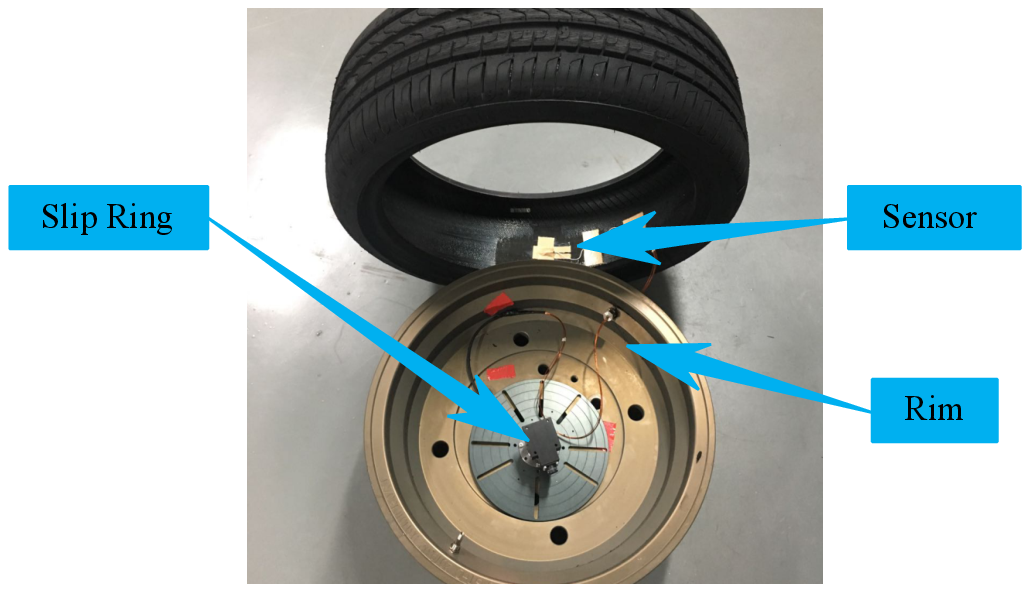}{b}
        \label{fig:second_sub}
    }
    \caption{(a) high speed type slip ring, (b)  its location on the metal}
    \label{figure_33}
\end{figure}

\begin{figure}[htb!]
  \begin{center}
  \includegraphics[width=2.7 in]{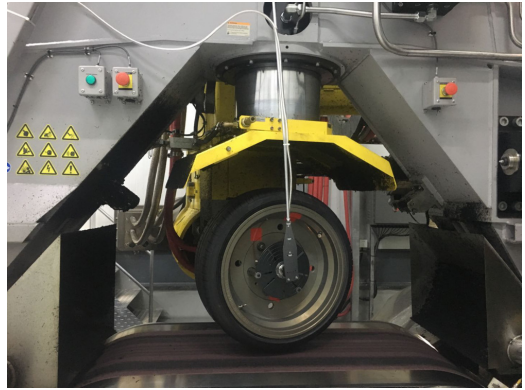}\\
  \caption{MTS tire testing system}\label{Figure_4}
  \end{center}
\end{figure}

The intelligent tire system and the Measure Test Simulate (MTS) Flat-Trac tire test platform are used in this study.~The intelligent tire system includes tri-axial acceleration sensor, slip ring, signal regulator, National Instrument (NI) data acquisition system (DAQ), as shown in Figure \ref{Figure_1}.~A tri-axial acceleration sensor is glued to the center of the tire inner liner, which is used to measure the longitudinal, lateral and vertical accelerations in the x, y and z directions of its body coordinate frame respectively, as depicted in Figure \ref{Figure_2}(a-b).~A slip-ring device is mounted to the rim to transmit the sensor signals from the rotating tire to the MTS Flat-Trac test platform, see Figure \ref{figure_33}(a-b) and \ref{Figure_4}.~The obtained acceleration signals are collected through the signal regulator and NI DAQ.~The signal debugger provides energy supply for the signal, while the NI acquisition system can adjust signal channel, sampling frequency and other settings to collect the acceleration signals.~The sampling rate  is 10kHz, which is high enough for the purpose of this study.


A set of tests by using Pirelli all-season tire (205/40R18) under free rolling, cornering and driving conditions with different vertical loads and velocities are used to train three different machine learning techniques to estimate the longitudinal, lateral and vertical forces (shown in Tables \ref{table_1}).~Typical acceleration signals in three directions are shown in Figure 5.
\begin{table}[htb!]
\centering
\caption{Test Conditions}
\begin{tabular}{p{2.5cm}p{4.5cm}}
 \hline
\centerline{~~~~~~~~~~~~~~~~~~~~~~~~~~~~~~~~~~~~~~~~~\textbf{Free Rolling}}
\\ \hline
\textbf{Test parameter}& \textbf{Value} \\ \hline
Velocity[kph] & 30/60/90
\\   \hline
 Pressure [kPa]
 & 220 \\ \hline
 Vertical load[N] & 2080/4160/6240/ \
 Triangular wave

 \\   \hline
 \centerline{~~~~~~~~~~~~~~~~~~~~~~~~~~~~~~~~~~~~~~~~\textbf{Cornering Condition}}
 \\ \hline
 Velocity[kph] & 30/60
 \\  \hline
 Pressure [kPa]
 & 220
 \\ \hline
 Vertical load[N] & 2080/4160/6240
 \\ \hline
 Slip angle [Deg.] & $ \pm 6\pm 5\pm 4\pm 3.5\pm 3\pm 2.5\pm 2\pm 1.5\pm 1$ \ Triangular wave
 \\ \hline
 \centerline{~~~~~~~~~~~~~~~~~~~~~~~~~~~~~~~~~~~~~~~~\textbf{Driving Condition}}
 \\ \hline
 Velocity[kph] & 30/60
\\   \hline
 Pressure [kPa]
 & 220 \\ \hline
 Vertical load[N] & 2080 \\ \hline
 Torque [N.m] & 207/218/343/400/442/526/565/650 \\ \hline
\label{table_1}
\end{tabular}
\end{table}

\begin{figure}
  \begin{center}
  \includegraphics[width=3.5in]{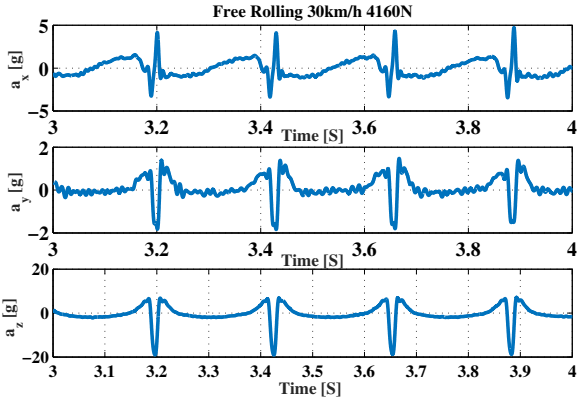}\\
  \caption{Typical acceleration signal}\label{Figure_5}
  \end{center}
\end{figure}
\section{Methodology}\label{section_3}
Not limited to several feature points, the complete acceleration information over the entire tire contact patch is used to estimate the tire forces in this research. The acceleration signals in x, y and z directions are inputs to the machine learning techniques, and the outputs are the longitudinal, lateral and vertical forces.

\subsection{Data pre-processing}
\begin{figure}
  \begin{center}
  \includegraphics[width=3.5in]{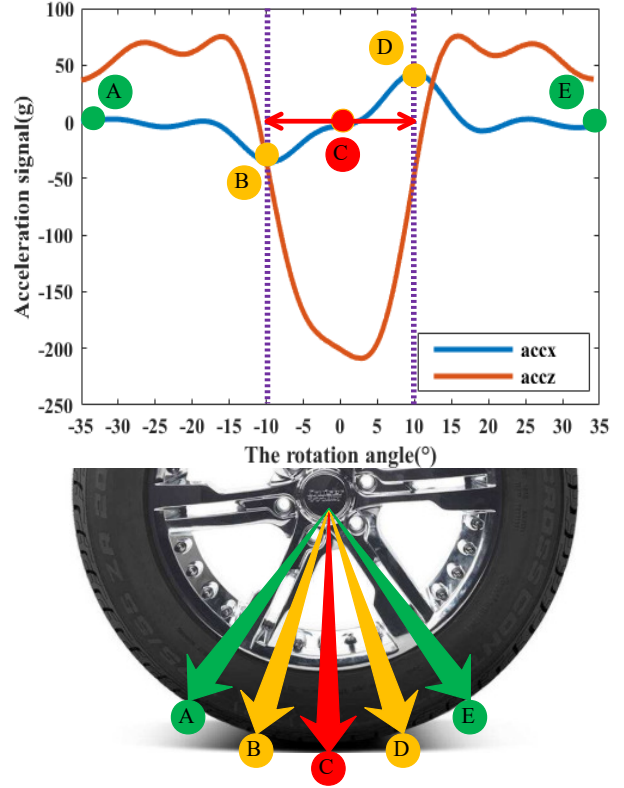}\\
  \caption{Processed acceleration data for machine learning}\label{Figure_6}
  \end{center}
\end{figure}

Data pre-processing transforms raw data and signals into the information used in machine learning techniques through a sequence of operations. It is a key step to the success of training process.~The objectives of data pre-processing include size reduction of the input space, smoother relationships, data normalization, noise mitigation, and feature extraction.~The main process will be briefly discussed in this part.
\begin{itemize}
\item Filtering and noise reduction: the sampling frequency of the acceleration signal is 10kHz, which is high enough for the purpose of tire force estimation. In this paper, the acceleration signal is filtered with 400Hz of cut-off frequency.~This is the main frequency induced by tire deformation.~The information in the higher frequency region could be used for other analysis, such as the micro-vibration between tire and road surface.
\item Identifying the contact patch: two peaks can be observed obviously in the circumferential direction when the accelerometer enters and leaves the contact patch.~Therefore, the data for every tire revolution can be extracted by using the encoder signal, and the acceleration peaks, which represent the contact patch region.~As shown in Figure 6, the points B and D are the starting  and ending positions of the contact patch, respectively.
\item  Transform the data in the contact patch from time series to spatial series:~only one acceleration sensor is used in order to save cost for the current study.~The history information is used when the tire with the attached sensor rolling through the contact zone.~To simplify the structure of neural network and make the number of features independent of the tire rolling speed, we extract time series data at fixed positions in the contact patch by using the encoder signal.~By identifying the contact patch in step 2, the center of contact patch C can be determined.~Finally,~one signal data per $0.5 ^\circ$ within $35^\circ$ rotation angle region extended from point C (from point A to E in the Figure \ref{Figure_6}) is collected for machine learning.
\item Data normalization: the Min-Max normalization is used for machine learning in this study.~The min-max normalization is a linear transformation, it can preserve all relationships of the data values exactly. It is given as the following equation:
\end{itemize}
\begin{equation}
    x_{norm}=\frac{x-x_{min}}{x_{max}-x_{min}}
\label{eq_1}
\end{equation}
where $x$ is the measured acceleration value, $x_{min}$ and $x_{max}$ are the minimum and maximum of the data.

\subsection{Neural Network}
An artificial neural network (ANN)–based formulation was developed to estimate the tire forces in terms of tire longitudinal, lateral and vertical acceleration.~Several advanced algorithms have been developed for neural networks learning. Broyden-Fletcher-Goldfarb-Shanno (BFGS), Levenberg-Marquardt, and conjugate gradients are classified as the well-known algorithms for training feedforward neural networks. Gradient descent methods (GDM) are also popular for supervised learning of neural networks.~One of the most efficient techniques based on GDM is batch back-propagation, which minimizes the error function implementing steepest descent method \cite{askari2019towards}.

\begin{figure}
  \begin{center}
  \includegraphics[width=3.5in]{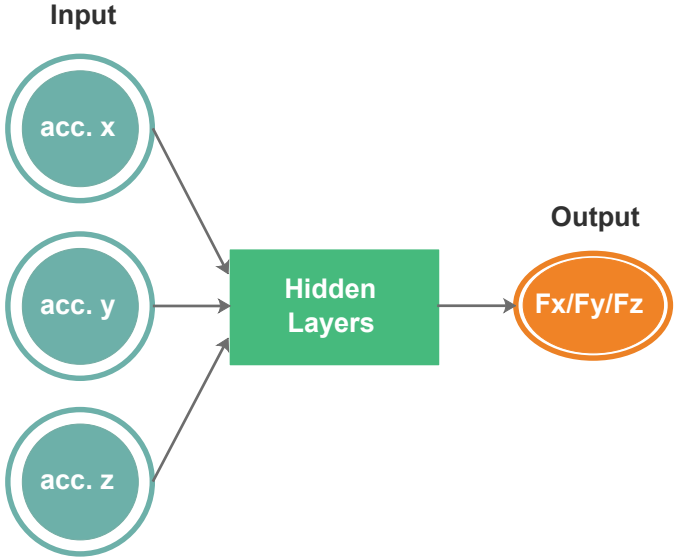}\\
  \caption{Layers of the implemented neural network for tire force estimation}\label{Figure_7}
  \end{center}
\end{figure}

Adaptive gradient-based algorithms have been also exploited for training neural networks.~They are considered as one of the trendiest algorithms for optimization, and also machine learning. Among different Adaptive gradient-based algorithms, Resilient backpropagation (Rprop) algorithm, which is the basis of neural networks in R, is considered as one of the best methods in terms of convergence speed, accuracy, and the robustness considering the learning parameters and rate. Using a sign-based technique, the Rprop algorithm updates the weights to avoid detrimental effects of derivative's magnitude on the updated weight.~As this method can effectively tackle the noisy error, they are eminently proper for implementation in hardware.~Therefore, the Rprop algorithm with logistic activation function and mean squared error (MSE) performance index is adopted in our research.~The structure of the neural network is shown in Figure \ref{Figure_7}, and three hidden layers (10-5-1) are considered for the neural network. For the $F_x$ and $F_z$ estimation, the longitudinal and vertical acceleration are put together as the inputs; for the $F_y$ estimation, acceleration signals in three directions are all used.

We divide the data into training, validation and test sets.~Training set is utilized to find the correlation between dependent and independent variables; validation dataset is used to tune model hyperparameters (i.e. the architecture); and the test set assesses the performance of the model.~We use 70$\%$, 15$\%$ and 15$\%$ of the dataset as training, validation and test set, respectively.~The assignment of the data to different sets is done using random sampling.
\section{Results and Discussions}
This section describes results of the tire force estimation by the machine learning method discussed in the previous sections.~They are trained based on the data sets described in Section~III, whereas the analysis presented here is purely performed on test datasets.~The performance of the designed algorithm is evaluated by the normalized root mean square error (NRMS), which is calculated by the following equation:
\begin{equation}
    NRMS=\frac{\sum_{i=1}^{N}(F_{measured}-{\hat{F}_{estimated}})^2}{max(\mid F_{measured} \mid)}
\end{equation}
where  $F_{measured}$,~${\hat{F}_{estimated}}$  and $N$ represent the measured signal, the estimated signal and the number of collected samples during the maneuver, respectively.
\subsection{Estimation of vertical forces}
\begin{figure}
  \begin{center}
  \includegraphics[width=3.5in]{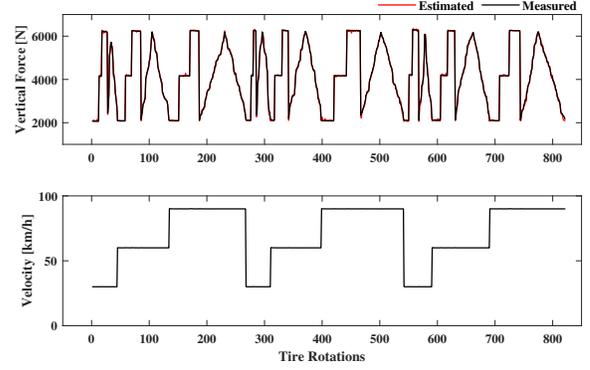}\\
  \caption{Estimation results for vertical tire forces under free rolling conditions}\label{Figure_8}
  \end{center}
\end{figure}

\begin{figure}
  \begin{center}
  \includegraphics[width=3.5in]{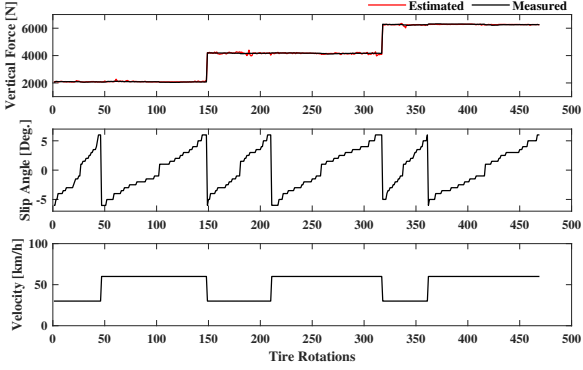}\\
  \caption{Estimation results for vertical tire forces under cornering conditions (step slip angle)}\label{Figure_9}
  \end{center}
\end{figure}

\begin{figure}
  \begin{center}
  \includegraphics[width=3.5in]{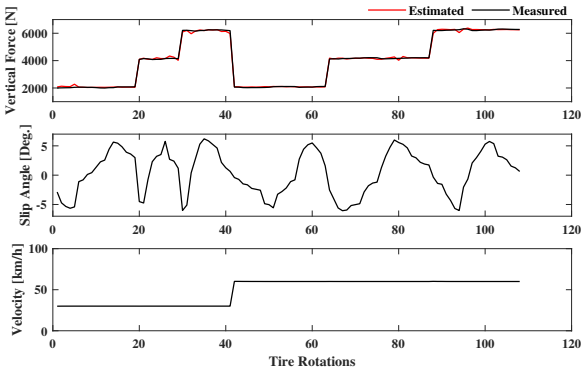}\\
  \caption{Estimation results for vertical tire forces under cornering conditions (triangle wave slip angle)}\label{Figure_10}
  \end{center}
\end{figure}

\begin{figure}
  \begin{center}
  \includegraphics[width=3.5in]{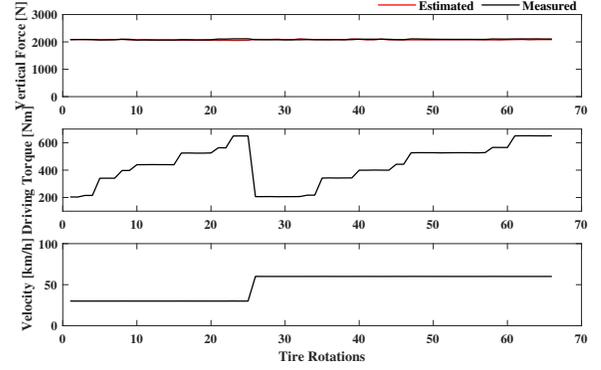}\\
  \caption{Estimation results for vertical tire forces under traction conditions (step driving torque)}\label{Figure_11}
  \end{center}
\end{figure}

\begin{figure}
  \begin{center}
  \includegraphics[width=3.5in]{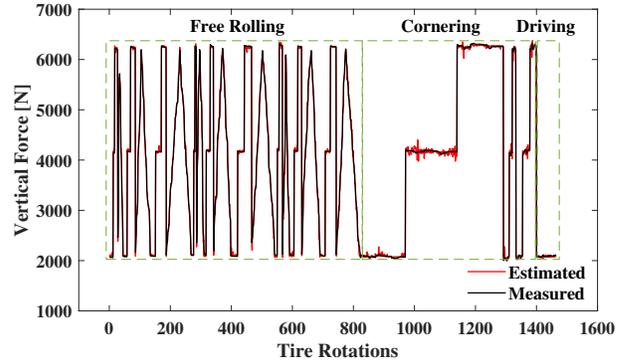}\\
  \caption{Estimation results for vertical tire forces (free rolling, cornering and driving conditions)}\label{Figure_12}
  \end{center}
\end{figure}

The performance of vertical load estimation is shown in Figs. \ref{Figure_8}-\ref{Figure_12}, which demonstrates an excellent agreement between the estimated and actual values of vertical forces for the testing scenarios. Fig. \ref{Figure_8} represents the data collected in free rolling conditions, which contain three days of data with repetitive experimental conditions using the experimental setup presented in Section \ref{expriment}. We applied step normal  and triangle wave loads to the intelligent tire from about 2kN to 6kN, which correspond to 40$\%$-120$\%$ of the tire load index. Three rolling speeds (30, 60 and 90 km/h) are used to include the influence of velocity on intelligent tire. From Fig. \ref{Figure_8}, it can be seen that the normal forces estimation under free rolling conditions exhibits high accuracy, and the adopted training network can effectively handle the influence of velocity changes.

For vehicle daily driving, it is important to include complex driving scenarios.~The vehicle will have a severe longitudinal/lateral load transfer and slip angle when the vehicle is driving or steering with large accelerations.~In Figs. \ref{Figure_9}-\ref{Figure_11}, the data is enlarged to contain cornering and driving conditions.~Figure \ref{Figure_9} shows the comparison results of tire vertical forces under cornering conditions, with step slip angle input from -6 to 6 degrees at 30kph and 60kph of speeds.~It is noted that the input data to neural network for tire vertical force estimation are the longitudinal and vertical accelerations, which means that the key features for determining tire vertical forces can be independent of tire slip angle and longitudinal velocity.~This is a good property for our trained network, since the slip angle of each tire and vehicle velocity are still a big challenge with current on-board sensors.

Figure \ref{Figure_10} shows the estimation results for vertical tire forces under cornering conditions with triangle wave slip angle input.~The slip angle changes from -6 degrees to 6 degrees, and the longitudinal velocities of tire are set to 30km/h and 60km/h.~The data sets used for testing of machine learning algorithm is extracted randomly from the overall data, so the shape of the slip angle input in Fig. \ref{Figure_10} seems a little irregular.~From both Fig. \ref{Figure_9}  and Fig. \ref{Figure_10}, we can see that the normal loads can be estimated accurately even though under cornering conditions with large slip angles.

To evaluate the performance of vertical force estimation under accelerating conditions, eight step driving torques are applied on the intelligent tire at 30km/h and 60km/h of speeds, as shown in Figure \ref{Figure_11}. It can be observed that the trained neural network is effective for estimating tire normal force under different driving conditions.~The overall estimation results of tire vertical forces are shown in Figure \ref{Figure_12}, containing the free rolling, cornering and driving conditions. As Figure \ref{Figure_12} shows, the estimated and real data are slightly different only in cornering region of the vertical loads.~The NRMS error is 0.81$\%$, which delineates the promising potential of machine learning technique for vertical load estimation of tire forces with a high accuracy.

\subsection{Estimation of lateral forces}
\begin{figure}[ht]
  \begin{center}
  \includegraphics[width=3.5in]{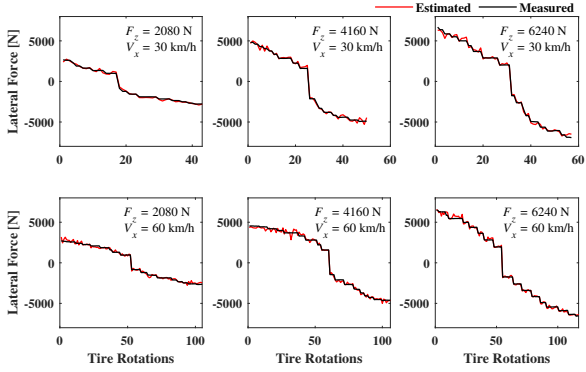}\\
  \caption{Estimation results for lateral tire forces (step input of slip angle)}\label{Figure_13}
  \end{center}
\end{figure}

\begin{figure}[ht]
  \begin{center}
  \includegraphics[width=3.5in]{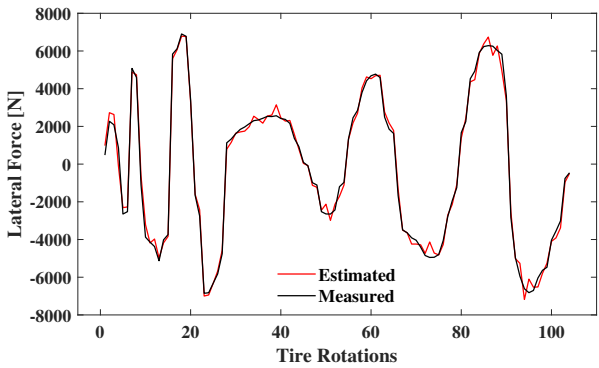}\\
  \caption{Estimation results for lateral tire forces (triangular wave input of slip angle)}\label{Figure_14}
  \end{center}
\end{figure}

\begin{figure}[ht]
  \begin{center}
  \includegraphics[width=3.5in]{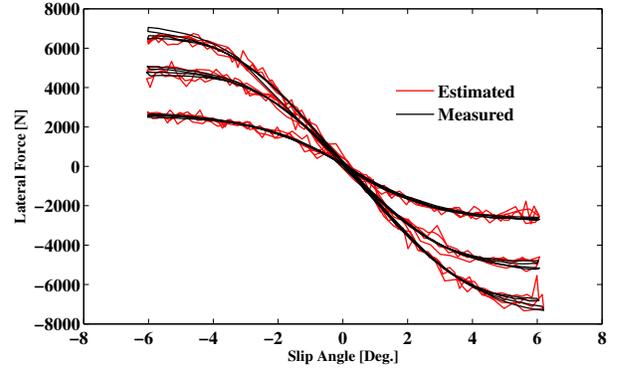}\\
  \caption{Tire lateral forces vs slip angle}\label{Figure_15}
  \end{center}
\end{figure}

Figures \ref{Figure_13}-\ref{Figure_15} show the lateral tire force estimation results under the triangular wave  and step inputs of slip angles.~The detailed test conditions can be found in Figs.\ref{Figure_9} and \ref{Figure_10}, which have shown the different slip angles, three loads and two velocities.~Figure \ref{Figure_13} demonstrates the comparison of measured and estimated lateral forces under the step input of slip angle, and Figure 14 represents the case of the triangle wave input of slip angles. It can be seen that the maximum lateral forces at each conditions are nearly equal to their corresponding limitation values, which is $\mu F_z$. It should be noted that the friction coefficient of the sandpaper on MTS test machine is about 1.1.

In our experiments, tires experience the maximum slip angle up to 6 degrees in which  the tire lateral force enters into the nonlinear region, as depicted by Figure 15. Figures 13-15 demonstrate that the machine learning technique can accurately predict the lateral forces of the tire.~Nevertheless, some differences exist in the regions of large lateral forces due to large slip angle (5-6 degrees).~The main source of these discrepancies may be the severe sliding and micro-vibration in the contact patch at large slip angles.~However, the overall performance of the method is satisfactory as its NRMS errors are lower than 5$\%$ (4.23$\%$). Comparing to previous works in the area of tire forces estimation with focus on lateral forces, the implemented machine learning technique shows a higher accuracy even in the case of large slip angles.

\subsection{Estimation of longitudinal forces}
\begin{figure}[htb]
  \begin{center}
  \includegraphics[width=3.5in]{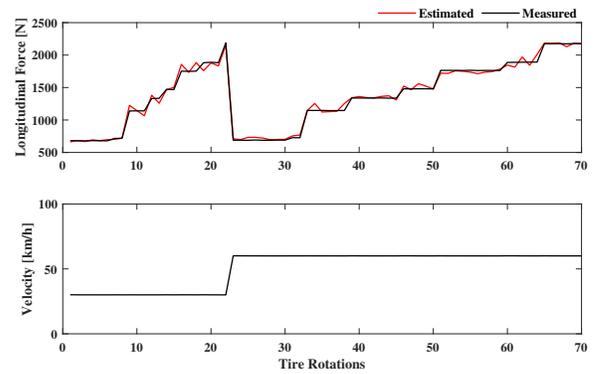}\\
  \caption{Estimation of tire longitudinal forces}\label{Figure_16}
  \end{center}
\end{figure}

In this section, we aim to focus on the potency of our trained machine learning algorithm for the estimation of tire longitudinal forces.~Figure \ref{Figure_16} compares the estimated and measured longitudinal forces.~The test conditions are set to step driving torque at 2880N load and two different velocities of 30 and 60 km/h. The results presented in this figure demonstrates the high accuracy of longitudinal tire force estimation based on the implemented machine learning technique.~With having larger data set, the accuracy of estimation can be even better for the longitudinal forces.~The NRMS error is about 2.89$\%$, which is satisfactory for tire force estimation.

\subsection{Cross validation}
\begin{figure}[htb]
  \begin{center}
  \includegraphics[width=3.5in]{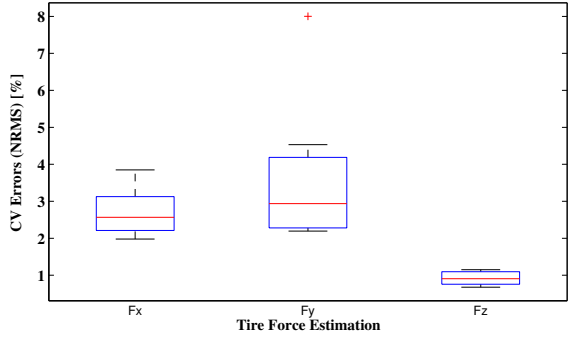}\\
  \caption{Boxplot of 10-fold cross-validation results}\label{Figure_17}
  \end{center}
\end{figure}

Cross-validation (CV) is a technique to assess the generalizability of a model to unseen data. The 10-fold cross-validation approach is adopted to reduce the bias associated with the random sampling of the training. In this cross-validation, the data set is divided into 10 portions, and 9 of them are utilized for training and 1 set is employed for testing.~The process is then repeated until all data are tested.~Fig. \ref{Figure_17} shows a box plot of validation results.~NRMS errors ranged from 1.98$\%$ to 3.85$\%$ for longitudinal force, 2.19 $\%$ to 4.53 $\%$ for lateral force, 0.68$\%$ to 1.15$\%$ for vertical force.~The average NRMS errors for tire forces estimation in three directions are 2.71$\%$, 3.55$\%$ and 0.92$\%$, respectively.~The CV results show that the trained model is reliable and has an accurate estimation for tire forces even under complex driving conditions.

\subsection{Comparison of the results with other machine learning methods}
As stated in the introduction section, Random Forest and Recurrent Neural Network are also commonly-used machine learning methods. Therefore, we used the same training and testing data for RF and RNN algorithms, and then compared with NN based on Rprop algorithm as shown in Figs. 18-20 for Fz, Fy and Fx separately. All training in this study were conducted on a laptop computer with Intel(R) Core(TM) i5-8250U CPU @ 1.60GHz and 16GB of DDR4 RAM. The hyper-parameters used for all models were assigned by “trial and error” approach. For Random Forest, the increase in the number of decision trees will improve the accuracy but the computational time required to train and test the system will increase. Most of the constant minimum error rate for each dataset for tire force estimation is achieved using 100 or 150 decision trees. We use a sgd optimizer and a mean squared error (MSE) for the RNN model. Three hidden layers (10-5-1) are also used, and the hyper-parameter configuration is 50, 10000 and 0.001 for the batch size, the number of epochs and the learning rate respectively.

Figures 18-20 show the tire forces curves of different machine learning methods, and Table \ref{Table2} summarizes the overall experiment results. It can be seen that Random Forest method converges faster, however, some obvious large errors exist for Random Forest, which may be owing to the limitation of extrapolation for RF when applied to regression problems. The prediction range of a Random Forest is bounded by the lowest and highest labels in the training data. This becomes a problem in situations, where the training and prediction inputs differ in the range and/or distributions. Thus, for extreme driving conditions, such as  large slip angles, severe noise of acceleration data may make the Random Forest perform poorly with testing data that is out of the range of the original training data. The RNN method shows a good capability in tire forces estimation similar or even better, in some cases, to NN based on Rprop algorithm. But it should be noted that the RNN algorithm is more complex and need more training time and more history information as input variables (10 times data used here). For vehicle safety control, it is necessary to keep the network as simple as possible in order to work in a real-time fashion. In addition, we try to avoid the network as temporal dependencies for direct tire force measurement/estimation technique of intelligent tire. Therefore, the neural network method based on Rprop algorithm is considered as the most effective approach for tire forces estimation in this paper.
\begin{figure}[htb]
  \begin{center}
  \includegraphics[width=3.5in]{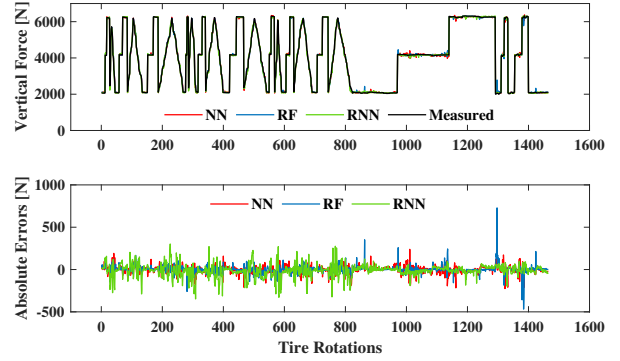}\\
  \caption{Comparison of estimated Fz with different ML methods}\label{Figure_18}
  \end{center}
\end{figure}

\begin{figure}[htb]
  \begin{center}
  \includegraphics[width=3.5in]{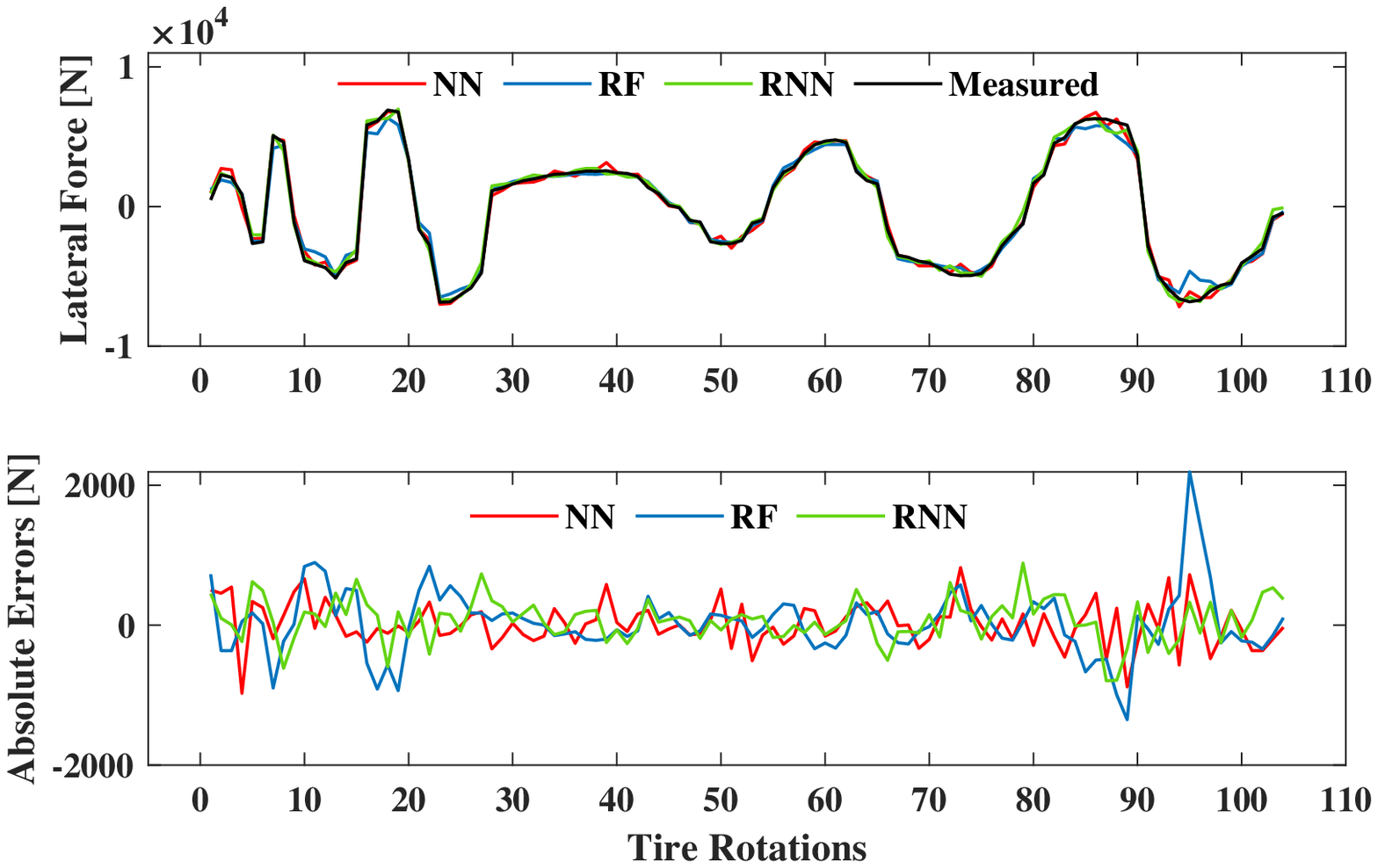}\\
  \caption{Comparison of estimated Fy with different ML methods}\label{Figure_19}
  \end{center}
\end{figure}

\begin{figure}[htb]
  \begin{center}
  \includegraphics[width=3.5in]{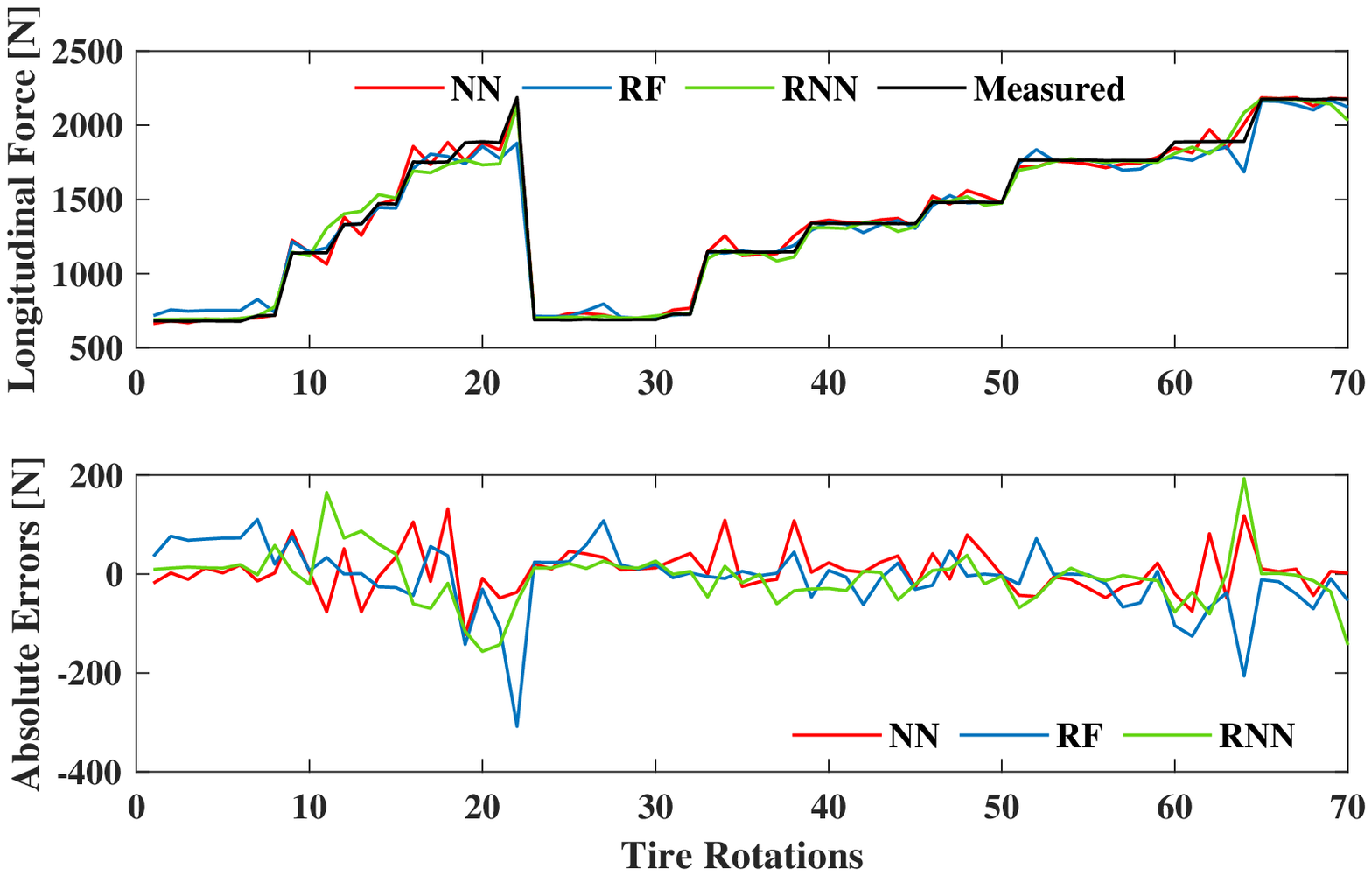}\\
  \caption{Comparison of estimated Fx with different ML methods}\label{Figure_20}
  \end{center}
\end{figure}

\begin{table}[htbp]
	\centering
	\caption{Summary of experiment results on different machine learning methods}\label{Table2}
	\newcommand{\tabincell}[2]{\begin{tabular}{@{}#1@{}}#2\end{tabular}}
	\begin{tabular}{c|c|c|c}
    \hline
	\tabincell{c}{Estimation \\variables} &
	\tabincell{c}{Estimation\\methods} &
	\tabincell{c}{Training time (s)/\\Data size} &
	\tabincell{c}{Testing dataset\\NRMS errors(\%)}
	\\
	\hline

    \multirow{3.7}{*}{Fz}
    & Neural network                            & 1499.99/6833         & 0.81                  \\
	\cline{2-4}
    & Random forest                             & 309.13/6833          & 0.95                  \\
    \cline{2-4}
	& \tabincell{c}{Recurrent \\neural network} & 6778.75/6833         & 1.42                  \\
	
	\hline	

	\multirow{3.7}{*}{Fy}
	& Neural network                            & 202.87/2713                     & 4.23                  \\
	\cline{2-4}
    & Random forest                             & 95.87/2713                      & 6.07                  \\
    \cline{2-4}
    & \tabincell{c}{Recurrent \\neural network} & 994.54/2713                     & 4.16                  \\

    \hline
     \multirow{3.7}{*}{Fx}
    & Neural network                            & 2.64/352                        & 2.89                  \\
    \cline{2-4}
    & Random forest                             & 1.94/352                        & 3.35                  \\
    \cline{2-4}
    & \tabincell{c}{Recurrent \\neural network} & 83.85/352                       & 3.67                  \\
    \hline
	\end{tabular}
\end{table}

\section{Conclusion}
In this paper, it was shown that the combination of machine learning and sensing system can be an ideal solution for the development of an intelligent tire.~Using an accelerometer attached to inner liner of a tire, we collected the acceleration data in x,y, and z directions plus tire real forces  in different testing scenarios.~The acceleration data and tire forces were then exploited for training a machine learning package with the aim of online prediction of tire forces.~A neural network was trained using the acceleration data and tire forces.~It was shown that tire forces could be accurately estimated using the trained network.~Based on the presented analysis, the trained machine learning package is capable of estimating tire vertical, lateral and longitudinal forces with the NRMS up to 1$\%$, 5$\%$, and 3$\%$, respectively.~In addition,a comparison was provided between NN based on Rprop algorithm, RNN,and RF.~It was shown that RNN and NN based on Rprop algorithm provides almost the same amount of accuracy in terms of tire forces estimation. However,NN based on Rprop algorithm is a better choice for the intelligent tire application owing to its simplicity in terms of real-time applications.~The presented research delineates the high potential of machine learning techniques for the realization of intelligent tire systems.

\bibliographystyle{IEEEtran}
\bibliography{IEEEabrv,Bibliography}
%


\begin{IEEEbiography}[{\includegraphics[width=1in,height=1.25in]{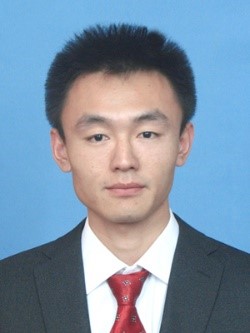}}]{Nan Xu}
received the Ph.D. degree in vehicle engineering from Jilin University, Changchun, China, in 2012. He is currently an associate professor at State Key Laboratory of Automotive Simulation and Control, Jilin University. His current research focuses on tire dynamics, intelligent tire, vehicle dynamics, stability control of electric vehicles and autonomous vehicles.
\end{IEEEbiography}
\begin{IEEEbiography}[{\includegraphics[width=1in,height=1.25in]{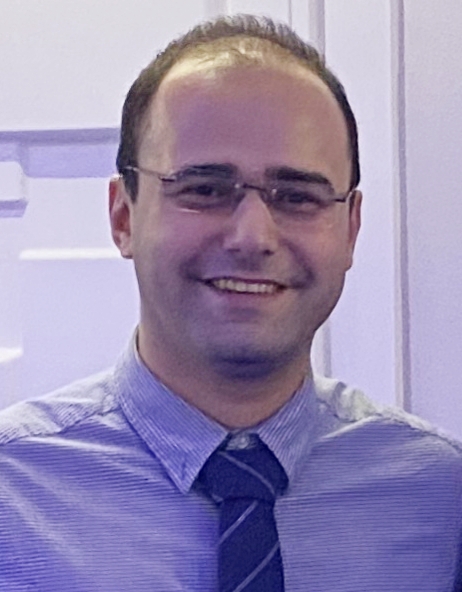}}]{Hassan Askari}
was born in Rasht, Iran and received his B. Sc. ,M.Sc. and PhD degrees from Iran University of Science and Technology, Tehran, Iran, University of Ontario Institute of Technology, Oshawa, Canada, and University of Waterloo, Waterloo, Canada in 2011, 2014, and 2019 respectively. He published more than 70 journal and conference papers in the areas of nonlinear vibrations, applied mathematics, nanogenerators, intelligent tires and self-powered sensors. He co-authored one book and one book chapter both published by Springer. He is an active reviewer for more than 40 journals and editorial board member of several scientific and international journals.  He is currently a Postdoctoral Fellow at the Department of Mechanical and Mechatronics Engineering at the University of Waterloo.
\end{IEEEbiography}
\begin{IEEEbiography}[{\includegraphics[width=1in,height=1.25in]{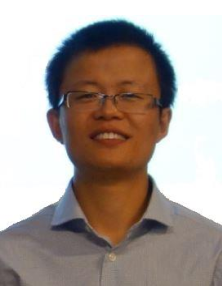}}]{Yanjun Huang}
 received his PhD degree from University of Waterloo in 2016. His research interest is mainly on the vehicle holistic control in terms of safety, energy-saving, and intelligence, including vehicle dynamics and control, HEV/EV optimization and control, motion planning and control of connected and autonomous vehicles, human-machine cooperative driving.He has published several books, over 60 papers in journals and conference; He is the recipient of IEEE Vehicular Technology Society 2019 Best Land Transportation Paper Award, the 2018 Best paper of Automotive Innovation, and top 10 most popular paper in
Journal Automobile Technology.~He is serving as associate editors and
editorial board member of IET Intelligent Transport System, SAE Int. J. of Commercial vehicles, Int. J. of Autonomous Vehicle system, etc
\end{IEEEbiography}

\begin{IEEEbiography}[{\includegraphics[width=1in,height=1.25in]{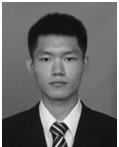}}]{Jianfeng Zhou}
 received his B.E. degree in automotive engineering in 2018 from Jinlin University, Changchun, China, where he is currently working toward the M.S. degree. His current research focuses on tire dynamics, intelligent tire and vehicle dynamics.
\end{IEEEbiography}
\begin{IEEEbiography}[{\includegraphics[width=1in,height=1.25in,clip,keepaspectratio]{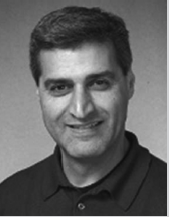}}]{Amir Khajepour}
is currently a Professor of
mechanical and mechatronics engineering with the
University of Waterloo, Waterloo, ON, Canada,
where he is also the Canada Research Chair in
mechatronic vehicle systems. He has developed an
extensive research program that applies his expertise in several key multidisciplinary areas. He is a
fellow of The Engineering Institute of Canada, The
American Society of Mechanical Engineers, and The
Canadian Society of Mechanical Engineering
\end{IEEEbiography}





\vfill


\end{document}